\documentclass[pra,twocolumn,showpacs]{revtex4}
\usepackage{epsfig,psfrag}
\usepackage{amsmath}
\allowdisplaybreaks

\begin{document}

\newcommand{\nfrac}[2]{\genfrac{}{}{0pt}{}{#1}{#2}} 

\title{Cooperative quantum jumps for three dipole-interacting atoms}
\author{Volker Hannstein} 
\author{Gerhard C.~Hegerfeldt}
\affiliation{Institut f\"ur Theoretische Physik, Universit\"at G\"ottingen,
  Friedrich-Hund-Platz~1, 37077 G\"ottingen, Germany}


\begin{abstract}
We investigate the effect of the dipole-dipole interaction on the
quantum jump statistics of three atoms. This is done for three-level
systems in a V configuration and in what may be called a D
configuration. The transition
rates between the four different intensity periods are calculated in
closed form. Cooperative effects are shown to increase by a factor of
two compared to two of either three-level systems. This results in
transition rates that are, for distances of about one wavelength of
the strong transition, up to $100\,\%$ higher than for independent
systems. In addition the double and triple jump rates are calculated
from the transition rates. In this case cooperative effects of up to
$170\,\%$ for distances of about one wavelength and still up to $15\,\%$
around 10 wavelengths are found. Nevertheless, for the parameters of
an experiment with Hg$^+$ ions the effects are negligible, in
agreement with the experimental data. For three Ba$^+$ ions this seems to
indicate that the large cooperative effects  observed experimentally
cannot be explained by the dipole-dipole interaction.

\end{abstract}

\maketitle

\section{Introduction}

Cooperative effects due to the dipole-dipole interaction between atoms
are of great importance in  
many fields, most recently in the study of possible quantum computers
based on trapped ions or atoms, and therefore they have attracted  considerable
interest in the literature \cite{refs:AdBeDaHe}.
A sensitive test for such cooperative effects can be provided by atoms
showing macroscopic light and dark periods in their
fluorescence. These can occur in a multilevel system if the electron
is essentially shelved in a metastable state, thereby causing the
photon emission to cease
\cite{ref:BeHe}.
Two or three such systems accordingly show three or four periods of different
intensity, namely one dark period and bright periods with once, twice,
or three times the intensity of a single system's bright period. The
dipole-dipole interaction may alter the statistics of these periods. 
In an as yet unexplained
experiment with two and three Ba$^+$ ions \cite{SaBlNeTo:86,Sa:86} a
large number of double and triple jumps, i.e., jumps by two or 
three intensity steps within a short resolution time, was observed,
exceeding by far the value expected for independent atoms.
The quantitative explanation of such large cooperative effects  for
distances of the order of 10 wavelengths of the strong transition has
been found difficult
\cite{HeNi:88,LeJa:87,LeJa:88,AgLaSo:88,LaLaJa:89,FuGo:92}. 
Experiments with other ions showed no observable cooperative effects
\cite{ThBaDhSeWi:92,BeRaTa:03}, in particular none were seen for
Hg$^+$ for a distance of about 15 wave lengths \cite{ItBeWi:88}. 
More recently effects similar to Ref.~\cite{SaBlNeTo:86} were found in
an experiment with Ca$^+$ ions \cite{BlReSeWe:99} in contrast to a comparable
experiment \cite{DoLuBaDoStStStSt:00}. A different method for
observing the dipole-dipole interaction of two V systems was proposed
in Ref.~\cite{SkoZaAgWeWa:01}. 

The effect of the dipole-dipole interaction for two V systems was
investigated numerically in Ref.~\cite{BeHe:99} and analytically in
Ref.~\cite{AdBeDaHe:01} and shown to be up to $30\,\%$
in the double jump rate compared to independent systems. However, the
systems used in the experimental setups of
Refs.~\cite{SaNeBlTo:86,SaBlNeTo:86,ItBeWi:88} cannot be described by
a V system so that a direct comparison between theory and experiment
was not possible. For this reason the present authors have
investigated cooperative effects for two other systems \cite{HaHe:03},
namely a D shaped system modeling the Hg$^+$ ions used in Ref.~\cite{ItBeWi:88}
and a four-level system modeling the Ba$^+$ ions of
Refs.~\cite{SaBlNeTo:86,Sa:86}. For two D systems cooperative effects
in the same order of magnitude as for the V systems were found for
ion distances of a few wavelengths of the laser-driven transition. For
larger distances practically no effects where found, in agreement with
the experiments \cite{ItBeWi:88} and with the 
results of Ref.~\cite{SkZaAgWeWa:01b}. In
contrast, only negligible effects for arbitrary ion-distances
were found for two of the four level-systems. Although this result
contradicts the findings of Refs.~\cite{SaBlNeTo:86,Sa:86} a direct
quantitative 
comparison with the experiments was not possible since explicit experimental
data were only provided for three Ba$^+$ ions. 

The aim of this paper is to narrow this gap by investigating {\em three}
dipole-interacting three-level systems in a V configuration and in a D
configuration (see Figs.~\ref{Vsystem} and \ref{Dsystem}),
respectively, and to compare the results with those for
two such systems. For three system this becomes much more
complicated since one has to deal with $729\times729$
matrices, and in order to do this we  use group theoretical
methods to exploit the symmetry of the problem.

We calculate the transition rates between
the different intensity periods for both systems. Cooperative effects
are found to increase by a factor of 2 in the first order terms in
the interaction parameter $C_3$ when compared to two of either
systems. This results in transition rates  up to $100\,\%$ higher than
the rates for independent systems. We 
also calculate the double and triple jump rates for both systems. Here
the cooperative effects are even larger.

A full description of the Ba$^+$ experiment
\cite{SaBlNeTo:86,Sa:86} would require the treatment of three of the
four-level systems of Ref.~\cite{HaHe:03}. However, here  we will restrict
ourselves to the three-level systems, since this reduces the
complexity of the calculation considerably. Also, the similarities
between the results for the D system and the four-level system pointed
out in Ref.~\cite{HaHe:03} seem to allow to draw conclusions on the cooperative
behavior of three four-level systems from the results presented
here. Namely, the increase of cooperative effects is not strong enough
to yield significant effects for three four-level systems.

Section \ref{3level} deals with the main
assumptions of the models. In Section \ref{transitionrates} the
methods for the calculation of the transition rates first for the V
systems and afterwards for the D system are explained. 
In Section \ref{results} the results of
the calculations are presented, namely the transition rates between
the different intensity periods. Finally in Section \ref{doublejump}
the double and triple jump rates are calculated from the transition
rates. The results are discussed and compared with those of two
three-level systems.
 
\section{Dipole-interacting three-level systems}
\label{3level}

\begin{figure}[b,t]
  \begin{center}  
      \epsfig{file=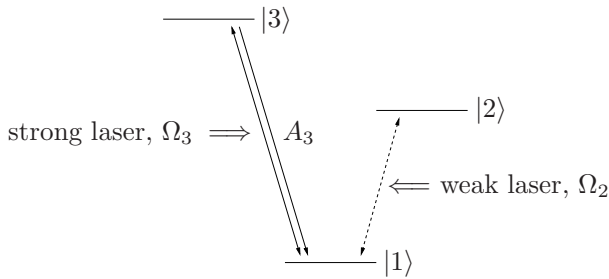,width=8cm}
    \caption{\label{Vsystem} Three-level system in V configuration.}
  \end{center}
\end{figure}
\begin{figure}[b]
  \epsfig{file=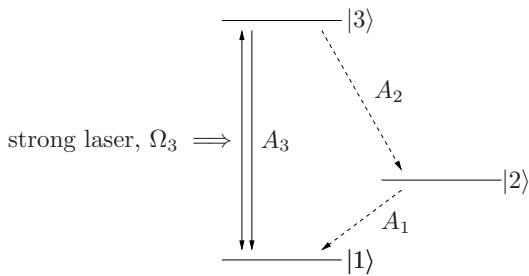,width=8cm}
  \caption{\label{Dsystem} Three-level system in D
      configuration with fast transitions (solid lines) and slow transitions
      (dashed lines).}
\end{figure}

In the following we investigate three dipole-interacting three-level
systems both in a V-type and in a D-type configuration as
shown in Figs.~\ref{Vsystem} and \ref{Dsystem}.    
For the V system the Rabi frequencies $\Omega_2$ and $\Omega_3$ and
the Einstein coefficient $A_3$ satisfy
\begin{equation}
\Omega_3,A_3 \gg \Omega_2
\end{equation}
so that the single system can show macroscopic light and dark periods.
The D system exhibits the same property if the condition
\begin{equation}
\Omega_3,A_3 \gg A_1,A_2
\end{equation}
for the Einstein coefficients and the Rabi frequency is fulfilled. We
assume the three atoms to be at fixed positions forming an
equilateral triangle, in agreement with the experimental
setups. Furthermore, for simplicity, the direction of the laser beams are
assumed to be perpendicular to the plane of this triangle.  

\begin{figure}[t!]
  \epsfig{file=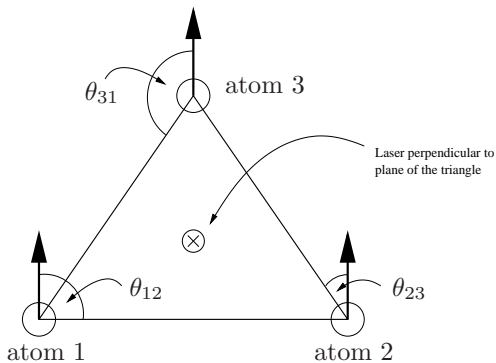,width=6.5cm}
  \caption{\label{atomsintrap} Geometry of the atoms in the trap. The
    arrows symbolize the dipole moments. In the picture the angles
    have the values $\theta_{12}=\pi/2$, $\theta_{23}=\pi/6$, and
    $\theta_{31}=5\pi/6$ leading to $\cos^2\theta_{12}=0$, and
    $\cos^2\theta_{23}=\cos^2\theta_{31}=3/4$.}
\end{figure}

The Bloch equation can be written in the form \cite{He:93}
\begin{equation}
\label{bloch}
  \dot{\rho}=-\frac{\text{i}}{\hbar}\left[H_{\text{cond}}\rho-\rho H_{\text{cond}}^{\dagger}\right]+ \mathcal{R}(\rho),
\end{equation}
where the conditional Hamiltonian $H_{\text{cond}}$ and the reset
operation $\mathcal{R}(\rho)$ for a general three-level system are
given by \cite{QJ,BeHe:98}
\begin{align}
\label{Hcond}
  H_{\text{cond}} & = 
  \sum_{i=1}^3\sum_{j=1}^3 \frac{\hbar}{2\text{i}}A_j S_{ij}^+S_{ij}^- + \sum_{i=1}^3 \sum_{j=2}^3 \frac{\hbar}{2} \left[
    \Omega_{j}S_{ij}^-  + \mbox{h.c.} \right] \nonumber \\* 
   & + \sum_{\nfrac{k,l=1}{k<l}}^3\sum_{j=1}^3
  \frac{\hbar}{2\text{i}} C_{kl}^{(j)}\left(S_{kj}^+ S_{lj}^- +
    S_{lj}^+S_{kj}^-\right)
\end{align}
and
\begin{align}
  \label{reset}
  \mathcal{R}(\rho) & = \sum_{i=1}^3\sum_{j=1}^3 A_j S_{ij}^-\rho
  S_{ij}^+ \nonumber \\
  & + \sum_{\nfrac{k,l=1}{k<l}}^3\sum_{j=1}^3
  \text{Re}\,C_{kl}^{(j)}\left(S_{kj}^-\rho S_{lj}^+ + S_{lj}^-\rho
    S_{kj}^+\right), 
\end{align}
with 
\begin{eqnarray}
  S_{i1}^+=|2\rangle_i {}_i\langle 1|, &&\quad S_{i2}^+=|3\rangle_i
  {}_i\langle 2|, \nonumber \\ \quad  S_{i3}^+=|3\rangle_i {}_i\langle
  1|, && \quad \text{and}
  \quad S_{ij}^-=S_{ij}^{+ \dagger}~.
\end{eqnarray}
Here,
\begin{eqnarray}
  C_{kl}^{(j)} & = &
  \frac{3A_j}{2}\text{e}^{\text{i}a_{kl}^{(j)}}\left[\frac{1}{\text{i}a_{kl}^{(j)}}(1 -\cos^2 \theta_{kl}) \right. \\ && \left. \hspace{1cm} {}
    + \left(\frac{1}{a_{kl}^{(j)2}}-\frac{1}{\text{i}a_{kl}^{(j)3}}\right)(1-3\cos^2 \theta_{kl})\right] \nonumber
\end{eqnarray}
is the coupling parameter which describes the dipole-dipole interaction
between atom $k$ and atom $l$ for the transition connected with the
Einstein coefficient $A_j$, with $\theta_{kl}$ being the angle between
the dipole moments and the line connecting the atoms. The dimensionless
parameter $a_{kl}^{(j)}=2\pi r_{kl}/\lambda_j$ is given by the
interatomic distance $r_{kl}$ multiplied by the wave number
$2\pi/\lambda_j$ of this transition. The detunings of the lasers are
taken as zero. 
By setting either $A_1=A_2=C_{kl}^{(1)}=C_{kl}^{(2)}=0$
or $\Omega_2=0$ in Eqs.~(\ref{Hcond}) and (\ref{reset}) the Hamiltonians and
reset states for the V systems and the D systems, respectively, are
obtained. 
For simplicity it would be preferable to have the same coupling
parameters for each pair of atoms (i.e., $C_{kl}^{(j)}\equiv
C_j$). This would 
be the case if the angle between the dipole moments and the line
connecting two atoms were the same for all pairs of atoms. However, the
arrangement of the atoms in the trap makes this impossible, as is
illustrated in Fig.~\ref{atomsintrap}. 
The atoms form an equilateral
triangle (i.e.,~$r_{kl}=r$) with the laser beams perpendicular to the
plane of this triangle and the dipole moments aligned by a magnetic field in a
direction in this plane. In this situation, the same value of the
coupling constants can only be achieved for two of the three possible
pairs of atoms. However, in spite
of this we will assume $C_{kl}^{(j)}\equiv C_j$  because this case
leads to maximal cooperative 
effects and can be seen as a limiting case for all other possible
configurations. The reset state can then be written as a sum of
density matrices of pure states
\begin{align}
\mathcal{R}(\rho) & = \sum_{j=1}^3\Big\{
  (A_j+2\text{Re}\,C_j)R_1^{(j)}\rho
  R_1^{(j)^{\scriptstyle \dagger}} \nonumber \\
  & \hspace{0.3cm} {}
  +(A_j-\text{Re}\,C_j)\Big[R_2^{(j)}\rho R_2^{(j)^{\scriptstyle
        \dagger}} +R_3^{(j)}\rho
    R_3^{(j)^{\scriptstyle \dagger}}\Big]\Big\},
\end{align}
with
\begin{align}
R_1^{(j)} & = \frac{1}{\sqrt{3}}\left(S_{1j}^- + S_{2j}^- +
  S_{3j}^-\right), \nonumber \\ 
R_2^{(j)} & =
\frac{1}{\sqrt{6}}\left(2S_{1j}^--S_{2j}^--S_{3j}^-\right), \\ 
R_3^{(j)} & = \frac{1}{\sqrt{2}}\left(S_{2j}^--S_{3j}^-\right) \nonumber.
\end{align}

In the case of two systems, it was convenient to use a Dicke basis,
i.e.,~a basis consisting of the symmetric and antisymmetric linear
combinations of the product states. Generally speaking, this means  
using a basis which is adapted with respect to the symmetry
group $S_2$ of permutations of two atoms. The symmetric and
antisymmetric states correspond to the irreducible representations of
this group. For three three-level systems, we therefore use a basis that is
adapted to the symmetry group $S_3$ of permutations of three
particles. On the subspace spanned by the product states with all
three atoms in 
different states the irreducible representations of the $S_3$ are the two one-dimensional representations mentioned above and another two equivalent
two-dimensional representations. This leads to the states
\begin{subequations}
\label{dicke}
\begin{align}
  |s_{123}\rangle & = 
  \frac{1}{\sqrt{6}}\big(|1\rangle|2\rangle|3\rangle
    +|2\rangle|3\rangle|1\rangle +|3\rangle|1\rangle|2\rangle 
  \nonumber \\* & \hspace{0.7cm} 
  {}+ |1\rangle|3\rangle|2\rangle + |2\rangle|1\rangle|3\rangle +
    |3\rangle|2\rangle|1\rangle \big), \\
  |a_{123}\rangle & = 
  \frac{1}{\sqrt{6}}\big(|1\rangle|2\rangle|3\rangle +|2\rangle|3\rangle|1\rangle
    +|3\rangle|1\rangle|2\rangle \nonumber \\* & \hspace{0.7cm}
  {}- |1\rangle|3\rangle|2\rangle -
    |2\rangle|1\rangle|3\rangle - |3\rangle|2\rangle|1\rangle \big), \\
  |b_{123}\rangle & = \frac{1}{\sqrt{12}}\big(2
    |1\rangle|2\rangle|3\rangle -|2\rangle|3\rangle|1\rangle
    -|3\rangle|1\rangle|2\rangle \nonumber \\* & \hspace{0.9cm} 
  {}+ 2|1\rangle|3\rangle|2\rangle -
    |2\rangle|1\rangle|3\rangle - |3\rangle|2\rangle|1\rangle \big), \\
  |c_{123}\rangle & =
  \frac{1}{2}\big(|2\rangle|3\rangle|1\rangle -
    |3\rangle|1\rangle|2\rangle \nonumber \\* & \hspace{0.7cm}
  {}- |2\rangle|1\rangle|3\rangle +
    |3\rangle|2\rangle|1\rangle \big), \\*
  |d_{123}\rangle & = \frac{1}{\sqrt{12}}\big(2|1\rangle|2\rangle|3\rangle
    -|2\rangle|3\rangle|1\rangle -|3\rangle|1\rangle|2\rangle
  \nonumber \\* & \hspace{0.9cm} 
  {}-2|1\rangle|3\rangle|2\rangle +|2\rangle|1\rangle|3\rangle
    +|3\rangle|2\rangle|1\rangle \big), \\ 
  |e_{123}\rangle & = \frac{1}{2}\big(|2\rangle|3\rangle|1\rangle -
    |3\rangle|1\rangle|2\rangle \nonumber \\ & \hspace{0.7cm} 
  {}+ |2\rangle|1\rangle|3\rangle - |3\rangle|2\rangle|1\rangle \big) 
\end{align}
\end{subequations}
in the case where all three atoms are in different states. For the
remaining states one then easily gets for $i,j=1,2,3$, $i \neq j$,
\begin{subequations}
  \begin{align}
    |s_{ijj}\rangle & =
    \frac{1}{\sqrt{3}}\big(|i\rangle|j\rangle|j\rangle + |j\rangle|j\rangle|i\rangle +
      |j\rangle|i\rangle|j\rangle \big), \\
    |b_{ijj}\rangle & =
    \frac{1}{\sqrt{6}}\big(2|i\rangle|j\rangle|j\rangle -
      |j\rangle|j\rangle|i\rangle - |j\rangle|i\rangle|j\rangle \big), \\
    |c_{ijj}\rangle & = \frac{1}{\sqrt{2}}\big(|j\rangle|j\rangle|i\rangle -
      |j\rangle|i\rangle|j\rangle\big) 
  \end{align}
\end{subequations}
if two atoms are in the same state and 
\begin{equation}
  |g\rangle = |1\rangle|1\rangle|1\rangle, \quad |e_2\rangle =
  |2\rangle|2\rangle|2\rangle, \quad |e_3\rangle = |3\rangle|3\rangle|3\rangle
\end{equation}
if all three atoms are in the same state.

\section{Transition Rates}
\label{transitionrates}
For the calculation of the transition rates, we carry over the methods
that have already been used for the description of two
dipole-interacting V systems and D systems, respectively
\cite{AdBeDaHe:01,HaHe:03}. 

For both types of systems, the configuration decouples
into four independent subspaces if one neglects the small parameters
(i.e., $\Omega_2=0$ for the V systems and $A_1=A_2=0$ for the D systems) 
\begin{subequations}
  \label{subspaces}
  \begin{align}
    \mathcal{S}_0 & = \{|e_2\rangle \}, \\*
    \mathcal{S}_1 & =  
    \{|s_{122}\rangle,|b_{122}\rangle,|c_{122}\rangle,|s_{322}\rangle,
    |b_{322}\rangle,|c_{322}\rangle\}, \\*
    \mathcal{S}_2 & = 
    \{|s_{211}\rangle,|b_{211}\rangle,|c_{211}\rangle,|s_{123}\rangle,
    |a_{123}\rangle, |b_{123}\rangle,\nonumber \\* & \hspace{0.65cm} 
    |c_{123}\rangle, |d_{123}\rangle,|e_{123}\rangle,|s_{233}\rangle,
    |b_{233}\rangle,|c_{233}\rangle\}, \\*
    \mathcal{S}_3 & = 
    \{|g\rangle,|s_{311}\rangle,|b_{311}\rangle,|c_{311}\rangle,\nonumber
    \\* & \hspace{0.65cm} 
    |s_{133}\rangle,|b_{133}\rangle,|c_{133},|e_3\rangle\} 
  \end{align}
\end{subequations}
in analogy to the case of two of either systems. In a period of
intensity $I_i$, the density matrix of the system is mostly in subspace
$\mathcal{S}_i$ \cite{BeHe:97}. The transition rates will thus be
calculated by using a density matrix in one particular subspace and
then  the rate of build-up of population in another subspace will be
determined. 

Taking a state $\rho_{0,i}$ in one of the subspaces $\mathcal{S}_i$ at a
time $t_0$ we calculate the state after a time $t_0 + \Delta t$
in perturbation theory with respect to the small parameters. The time interval
used here should be long in comparison to the mean time between the
emission of two photons but short in comparison to the length of the
intensity periods,
\begin{eqnarray}
A_3^{-1},\Omega_3^{-1} & \ll & \Delta t \ll \Omega_2^{-1} \quad \text{(V
  system)}, \nonumber \\
A_3^{-1},\Omega_3^{-1} & \ll & \Delta t \ll A_1^{-1},A_2^{-1} \quad
\text{(D system)}~.
\end{eqnarray}
For the calculation the Bloch equation is written in a Liouvillean form
\begin{equation}
\dot{\rho}=\mathcal{L}\rho =
\{\mathcal{L}_0(A_3,C_3,\Omega_3)+\mathcal{L}_1\} \rho~,
\end{equation}
where $\mathcal{L}_1$ serves as the perturbation depending on
$\Omega_2$ or $A_1,A_2,C_1$, and $C_2$, respectively.
We then get \cite{AdBeDaHe:01}
\begin{equation}
  \label{rhotdeltat}
  \rho(t_0+ \Delta t;\rho_{\text{ss},i}) = \rho_{\text{ss},i}+ \int_0^{\Delta t}\text{d}\,\tau e^{\mathcal{L}_0\tau}\mathcal{L}_1\rho_{\text{ss},i},
  \end{equation}
where $\rho_{\text{ss},i}$ is the quasisteady state in subsystem
$\mathcal{S}_i$. As a Liouvillean of Bloch equations, $\mathcal{L}_0$
has an eigenvalue $0$ corresponding to the quasisteady states. The
other eigenvalues have negative real parts of the order of $\Omega_3$ and
$A_3$. While $\mathcal{L}_1\rho_{\text{ss},i}$ is a
superposition of just the eigenstates for nonzero eigenvalues of
$\mathcal{L}_0$ in the case of three V systems this is not true for
three D systems, which makes it necessary to discuss the two cases separately.

\subsection{Three V systems}
For the V systems, $\mathcal{L}_1\rho_{\text{ss},i}$ consists only
of coherences between the subspace $\mathcal{S}_i$ and the
neighbouring subspaces, since $\mathcal{L}_1$ describes the coupling
due to the weak laser (with Rabi frequency $\Omega_2$) in this
case. The zero-eigenvalue subspace of $\mathcal{L}_0$, on the other hand,
is spanned by the quasisteady states $\rho_{\text{ss},i}$. Therefore, $\mathcal{L}_1\rho_{\text{ss},i}$ has no components in
the zero eigenvalue subspace of $\mathcal{L}_0$ in the case of V
systems. The other eigenvalues all have negative
real parts of the order of $A_3$ and $\Omega_3$. Therefore the integrand in
Eq.~(\ref{rhotdeltat}) is rapidly damped which allows us to extend the
upper integration limit to infinity. This yields
\begin{equation}
\label{rhotdeltat2}
\rho(t_0+ \Delta t;\rho_{\text{ss},i}) = \rho_{\text{ss},i} + (\epsilon-\mathcal{L}_0)^{-1}\mathcal{L}_1\rho_{\text{ss},i}~,
\end{equation}   
independent of $\Delta t$ \cite{AdBeDaHe:01}.

From the Bloch equations (\ref{bloch}) we get the exact relations 
\begin{subequations}
\label{rhodot}
\begin{widetext}
\begin{eqnarray}
  \frac{\text{d}}{\text{d}t}\langle e_2|\rho|e_2\rangle & = & \sqrt{3}\,\Omega_2 \text{Im}\,
  \langle s_{122}|\rho |e_2\rangle~, \\
  \frac{\text{d}}{\text{d} t}\sum_{x_i \in \mathcal{S}_1}\langle x_i|\rho|x_i\rangle
  & = & \Omega_3\text{Im}\, \bigg[2\langle s_{112}|\rho|s_{122}\rangle-\langle
  b_{112}|\rho|b_{122}\rangle -\langle c_{112}|\rho|c_{122}\rangle-
  \sqrt{3}\langle s_{122}|\rho|e_2\rangle + \sqrt{2}\langle
  s_{123}|\rho|s_{223}\rangle \nonumber \\ 
  && {}- \frac{1}{\sqrt{2}}\left(\langle
      b_{123}|\rho|b_{223}\rangle + \langle
      c_{123}|\rho|c_{223}\rangle \right) +
    \sqrt{\frac{3}{2}}\left(\langle d_{123}|\rho|c_{223}\rangle -
      \langle e_{123}|\rho|b_{223}\rangle \right)\bigg] - \frac{\text{d}}{\text{d}
    t}\langle e_2|\rho|e_2\rangle~, \\
  \frac{\text{d}}{\text{d} t}\sum_{x_i \in \mathcal{S}_2}\langle x_i|\rho|x_i\rangle
  & = & {}- \Omega_3\text{Im}\, \bigg[2\langle s_{112}|\rho|s_{122}\rangle-\langle
    b_{112}|\rho|b_{122}\rangle -\langle c_{112}|\rho|c_{122}\rangle -
    \sqrt{3}\langle s_{122}|\rho|e_2\rangle + \sqrt{2}\langle
    s_{123}|\rho|s_{223}\rangle \\
  && {}- \frac{1}{\sqrt{2}}\left(\langle
      b_{123}|\rho|b_{223}\rangle + \langle
      c_{123}|\rho|c_{223}\rangle \right)+
    \sqrt{\frac{3}{2}}\left(\langle d_{123}|\rho|c_{223}\rangle -
      \langle e_{123}|\rho|b_{223}\rangle \right)\bigg] - \frac{\text{d}}{\text{d}
    t} \sum_{x_i \in \mathcal{S}_3}\langle x_i|\rho|x_i\rangle~, \nonumber \\
  \frac{\text{d}}{\text{d} t}\sum_{x_i \in \mathcal{S}_3}\langle x_i|\rho|x_i\rangle
  & = & \Omega_2 \text{Im}\, \bigg[\frac{1}{\sqrt{2}}(\langle
  b_{113}|\rho|b_{123}\rangle + \langle c_{311}|\rho|c_{123}\rangle) + 
  \sqrt{\frac{3}{2}}(\langle b_{311}|\rho|e_{123}\rangle -
  \langle c_{311}|\rho|d_{123}\rangle) - \sqrt{3}\langle
  g|\rho|s_{211}\rangle  \nonumber \\ 
  & & {}- \sqrt{2}\langle s_{311}|\rho|s_{233}\rangle
    -(\langle s_{133}|\rho|s_{233}\rangle+\langle
    b_{133}|\rho|b_{233}\rangle + \langle c_{133}|\rho|c_{233}\rangle) \bigg].
\end{eqnarray}
\end{widetext}
\end{subequations}
Together with Eq.~(\ref{rhotdeltat2}) this allows us to calculate the
transition rates as
\begin{equation}
\label{pijcalc}
p_{ij}=\frac{\text{d}}{\text{d} t}\sum_{x_k \in \mathcal{S}_j}\langle
x_k|\rho|x_k\rangle \Big|_{\rho=\rho(t_0+ \Delta t;\rho_{\text{ss},i})} .
\end{equation} 
Note that $p_{ij}=0$ for $|i-j|\geq 2$ so that no {\em direct},
i.e., instantaneous, double jumps occur. 

\subsection{Three D systems}
In the case of D systems, $\mathcal{L}_1$ describes spontaneous
emission due to the Einstein coefficients $A_1$ and $A_2$. Therefore
$\mathcal{L}_1\rho_{\text{ss},i}$ consists of density matrix elements
$\langle x_i|\rho|x_j\rangle$ where both states $|x_i\rangle$ and
$|x_j\rangle$ lie in the same subspace $\mathcal{S}_i$. It is thus a
superposition of eigenstates of $\mathcal{L}_0$ with zero as well as nonzero
eigenvalues. We write
\begin{equation}
  \label{L1rho0}
  \mathcal{L}_1\rho_{\text{ss},i}=\sum_{j=0}^3 \alpha_{ij}\rho_{\text{ss},j} + \tilde{\rho},
\end{equation}
where $\tilde{\rho}$ contains the contributions from the eigenstates
for nonzero eigenvalues of $\mathcal{L}_0$. The coefficients
$\alpha_{ij}$ are calculated by means of the dual eigenstates
$\rho_{\text{ss}}^i$ \cite{HaHe:03},
\begin{equation}
\label{alpha}
  \alpha_{ij}=\text{Tr}(\rho_{\text{ss}}^{j\dagger}\mathcal{L}_1\rho_{\text{ss},i}).
\end{equation}
Inserting Eq.~(\ref{L1rho0}) into Eq.~(\ref{rhotdeltat}) one obtains
\begin{eqnarray}
  \label{rhotdeltat3}
  \rho(t_0+\Delta t) = \rho_{\text{ss},i} + \sum_{j=0}^3
  \alpha_{ij}\rho_{\text{ss},j}\Delta t +
  (\epsilon-\mathcal{L}_0)^{-1}\tilde{\rho}~.  
\end{eqnarray}
The last term is much smaller than the preceding term and can be
neglected \cite{HaHe:03}. The coefficients $\alpha_{ij}$ can then be
interpreted as the transition rates between the subspaces
$\mathcal{S}_i$ and $\mathcal{S}_j$,  
\begin{equation}
\label{pij}
p_{ij}=\alpha_{ij}.
\end{equation} 

\subsection{Group theory}
\label{group} 
For the calculation of the transition rates for both V systems and
D systems it is necessary to calculate the quasisteady states
$\rho_{\text{ss},i}$, i.e.,~to solve the linear equation
\begin{equation}
\label{rhoss}
\mathcal{L}_0\rho_{\text{ss}}=0.
\end{equation}
In addition, for V systems the first order term 
\[
\rho^{(1)}_i =
(\epsilon-\mathcal{L}_0)^{-1}\mathcal{L}_1\rho_{\text{ss},i}
\] 
of Eq.~(\ref{rhotdeltat2}) must be calculated, which was done by
solving 
\begin{equation}
\label{rho1}
\mathcal{L}_0\rho^{(1)}_i = \mathcal{L}_1\rho_{\text{ss},i}~.
\end{equation}
Eqs.~(\ref{rhoss}) and (\ref{rho1}) are linear equations for the 729
matrix elements of $\rho_{\text{ss},i}$ and $\rho^{(1)}_i$,
respectively. Luckily there are two different properties of
$\mathcal{L}_0$ that make it possible to restrict these equations to
smaller subspaces, which reduces the calculation effort considerably.
First, $\mathcal{L}_0$ is independent of the small parameters ($A_1$,
$A_2$ or $\Omega_2$), which means that there is no coupling between
the four subspaces of Eq.~(\ref{subspaces}). Thus there exist 16
subspaces $\mathcal{R}_{i,j}$,  each
consisting of the density matrix elements
\begin{equation}
\langle x_i|\rho|y_j\rangle \quad \text{with} \quad |x_i\rangle \in
\mathcal{S}_i \quad \text{and} \quad |y_j\rangle \in \mathcal{S}_j~, 
\end{equation}
respectively, which are invariant with respect to $\mathcal{L}_0$.
In addition the conditional Hamiltonian $H_{\text{cond}}$ and the reset
state $\mathcal{R}(\rho)$ and therefore also $\mathcal{L}_0$ are
invariant under the exchange of atoms, as can be seen from
Eqs.~(\ref{Hcond}) and (\ref{reset}). Hence subspaces which consist of
all density matrix elements which belong to a particular irreducible
representation of $\mathcal{S}_3$ are also invariant with respect to
$\mathcal{L}_0$. Since the density matrix elements form a
representation of $\mathcal{S}_3$ which is a tensor product of twice
the representation spanned by the Dicke basis of Eq.~(\ref{dicke}) the
new irreducible representations are easily found. The density matrix elements
\begin{align}
  |s_\alpha\rangle\langle s_\beta|, \quad |a_\alpha\rangle\langle a_\beta|, & 
  \nonumber \\*
  \,\frac{1}{2}\left(|b_\alpha\rangle\langle
    b_\beta|+|c_\alpha\rangle\langle c_\beta| \right), & \quad
  \frac{1}{2}\left(|d_\alpha\rangle\langle
    d_\beta|+|e_\alpha\rangle\langle e_\beta| \right), \\*
  \frac{1}{2}\left(|b_\alpha\rangle\langle
    e_\beta|-|c_\alpha\rangle\langle d_\beta| \right), & \quad  
  \frac{1}{2}\left(|e_\alpha\rangle\langle
    b_\beta|-|d_\alpha\rangle\langle c_\beta| \right)\, \nonumber
\end{align} 
belong to the symmetric representation, the elements
\begin{align}
  |s_\alpha\rangle\langle a_\beta|, \quad |a_\alpha\rangle\langle s_\beta|, &
  \nonumber \\
  \,\frac{1}{2}\left(|b_\alpha\rangle\langle
    c_\beta| - |c_\alpha\rangle\langle b_\beta| \right),& \quad
  \frac{1}{2}\left(|d_\alpha\rangle\langle
    e_\beta|-|e_\alpha\rangle\langle d_\beta| \right), \\
  \frac{1}{2}\left(|b_\alpha\rangle\langle
    d_\beta|+|c_\alpha\rangle\langle e_\beta| \right),& \quad
  \frac{1}{2}\left(|d_\alpha\rangle\langle 
    b_\beta|+|e_\alpha\rangle\langle c_\beta| \right)\, \nonumber  
\end{align}
belong to the antisymmetric representation, and  the remaining 24
possible linear combinations form two-dimensional representations.
Here $\alpha$ and $\beta$ are one of the subscripts of the Dicke
states. By transforming the Liouvillean $\mathcal{L}_0$ into this new
basis each of the 16 invariant subspaces $\mathcal{R}_{i,j}$ is in
itself decomposed into three invariant subspaces connected to the
elements belonging to the symmetric, antisymmetric, and
two-dimensional representations, respectively. For the calculation of
both the quasisteady states $\rho_{\text{ss},i}$ and the
transition rates for the V systems only the symmetric subspaces are
needed. For the latter this can be seen from Eq.~(\ref{rhodot}). With
these two simplifications the dimension of the linear system of
equations needed for the calculation reduces considerably (namely to a
maximum of 20 for the calculation of $p_{23}$ and $p_{32}$).

\section{results}
\label{results}
The transition rates for the V systems can now be calculated according
to Eqs.~(\ref{pijcalc}), (\ref{rhotdeltat}), and (\ref{rhodot}). The
result is
\begin{subequations}
\label{pijV}
\begin{align}
p_{01} & = 3\frac{A_3\Omega_2^2}{\Omega_3^2}~, \\
p_{10} & = \frac{A_3^3\Omega_2^2}{\Omega_3^2(A_3^2+2\Omega_3^2)}~, \\
p_{12} & = 2\frac{A_3\Omega_2^2}{\Omega_3^2}\left[1 + 2\text{Re}\,
  C_3\frac{A_3}{A_3^2+2\Omega_3^2}\right], \\
p_{21} & =   
2\frac{A_3^3\Omega_2^2}{\Omega_3^2(A_3^2+2\Omega_3^2)}\left[1+ 2\text{Re}\,
  C_3\frac{A_3(A_3^2+4\Omega_3^2)}{(A_3^2+2\Omega_3^2)^2} \right], \\
p_{23} & = \frac{A_3\Omega_2^2}{\Omega_3^2}\left[1 + 4\text{Re}\,
  C_3\frac{A_3}{A_3^2+2\Omega_3^2}\right], \\
p_{32}  & = 3\frac{A_3^3\Omega_2^2}{\Omega_3^2(A_3^2+2\Omega_3^2)}\left[1+ 4\text{Re}\,
  C_3\frac{A_3(A_3^2+4\Omega_3^2)}{(A_3^2+2\Omega_3^2)^2} \right]  
\end{align}
\end{subequations}
to first order in $C_3$. 
\begin{figure*}[t!]
  \begin{minipage}[t]{0.47\textwidth}
    \psfrag{p32}{$p_{32} [\text{s}^{-1}]$}
    \psfrag{r/lambda3}{$r [\lambda_3]$}
    \epsfig{file=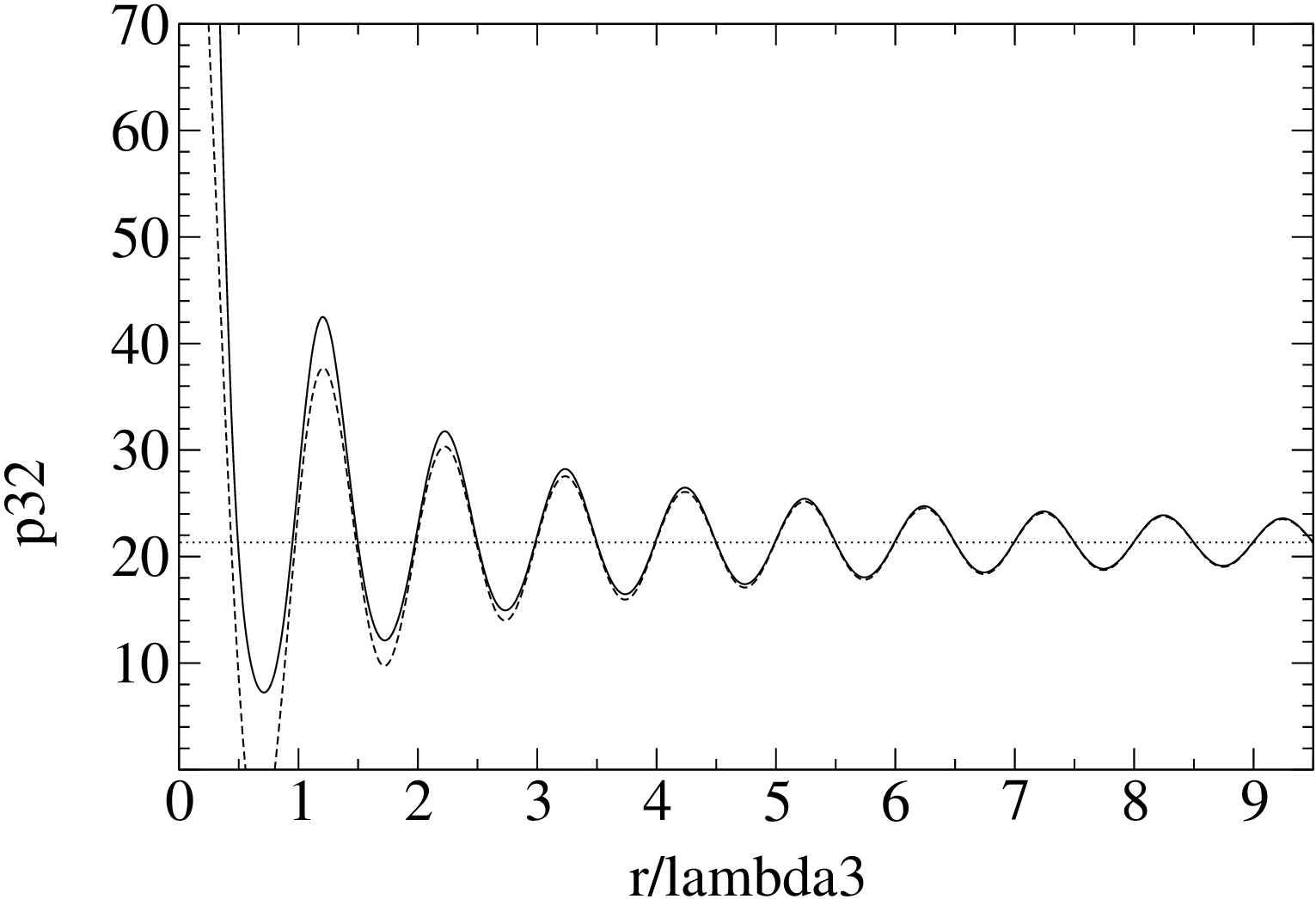, width=7.5cm,height=5cm}
    \caption{\label{3Vp32} Transition rate $p_{32}$ for three
      dipole-interacting V systems plotted versus the interatomic
      distance $r$ in units of the wavelength $\lambda_3$ of the
      strong transition. Solid line: $p_{32}$ up to second
      order in $C_3$. Dashed line: first order. Dotted line:
      independent systems. Parameter values are $A_3=2\cdot
      10^8\,\text{s}^{-1}$, $\Omega_3=5\cdot10^7\,\text{s}^{-1}$, and
      $\Omega_2=10^4\,\text{s}^{-1}$.} 
\end{minipage}
\hfill
\begin{minipage}[t]{0.47\textwidth}
\psfrag{p32}{$p_{32} [\text{s}^{-1}]$}
  \psfrag{r/lambda3}{$r [\lambda_3]$}
   \epsfig{file=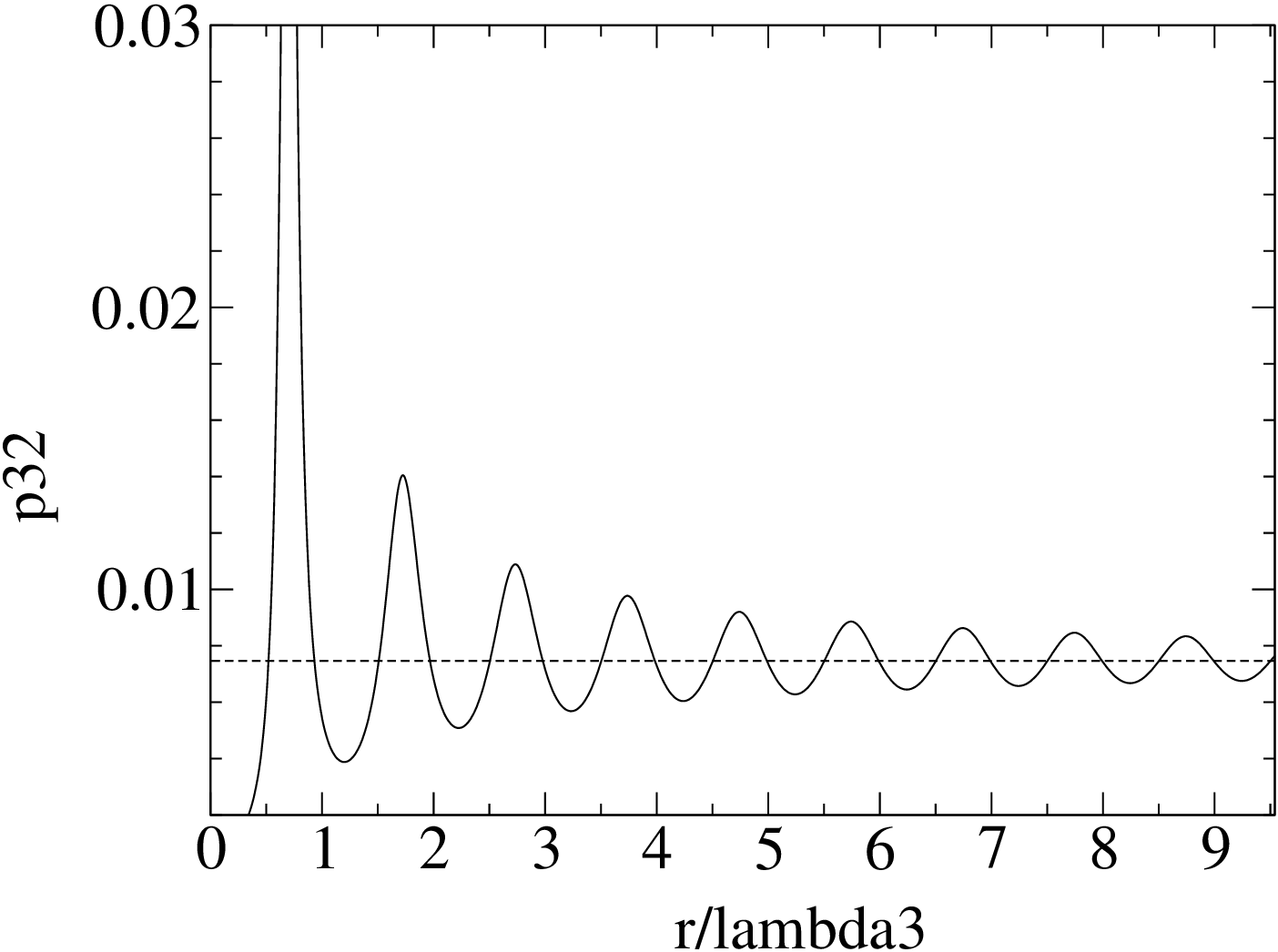, width=7.5cm,height=5cm}
   \caption{\label{3Dp32}  Transition rate $p_{32}$ for three
     dipole-interacting D systems. Dashed line: independent systems. 
     Parameter values are $A_1=1\,\text{s}^{-1}$,
     $A_2=1\,\text{s}^{-1}$, $A_3=2\cdot 10^8 \,\text{s}^{-1}$, and
     $\Omega_3=10^7\,\text{s}^{-1}$.} 
\end{minipage}
\end{figure*}
While for $p_{01}$ and $p_{10}$ this is also
the exact result to all orders, the higher order terms for the
other four transitions are too complicated to be given here.  
The zeroth order terms in
Eqs.~(\ref{pijV}) are those one would expect for independent atoms
(namely the rates $p_{10}$ and $p_{01}$ for single V system multiplied
by a factor 1, 2, or 3). For the first order terms it is important to
note that the single systems interact via $C_3$ only if they are in a
light period. Therefore the rates $p_{01}$ and $p_{10}$ are
independent of $C_3$ while $p_{12}$ and $p_{21}$ have the same first
order term as the corresponding rates for two V systems (in the intensity
period $I_2$ the three V systems behave like two V systems in the period $I_2$
plus an additional noninteracting  system). 
In the rates $p_{23}$ and $p_{32}$ the first order term is just twice the first
order term of $p_{21}$ and $p_{12}$. This surprising property is due
to the simplicity of the quasisteady state $\rho_{\text{ss},3}$,
namely all
diagonal elements of this state have the same first order dependence.
Fig.~\ref{3Vp32} shows the transition rate $p_{32}$ for three V
systems to first and to second order in $C_3$. The first order rate
becomes negative for distances of about one-half to three quarters of
a wavelength of the strong transition. By looking at the second order
rate one can see that is an artefact of the approximation. The rate
with the dipole interaction included shows deviations of up to $100\,\%$
from the rate for 
noninteracting atoms for distances of somewhat more than a wavelength
$\lambda_3$.

By use of Eqs.~(\ref{rhoss}), (\ref{alpha}) and (\ref{pij}) the
transition rates for three dipole interacting D systems were also
calculated, with the result
\begin{subequations}
\begin{align}
  p_{01} & = 3A_1, \quad
  p_{12} = 2A_1, \quad
  p_{23} = A_1,  \\
  p_{10} & = \frac{A_2\Omega_3^2}{A_3^2+2\Omega_3^2}  
\end{align}
and
\begin{widetext}
\begin{align}
  \label{2Dp21}
  p_{21} & = 2\frac{A_2\Omega_3^2(A_3^2+2\Omega_3^2)}{(A_3^2 +
    2\Omega_3^2)^2 + A^2_3(|C_3|^2+
    2A_3\text{Re}\,C_3)}   
  = \frac{2A_2\Omega_3^2}{A_3^2+2\Omega_3^2}\left[1
    - 2\text{Re}\,C_3\frac{A_3^3}{(A_3^2 +
      2\Omega_3^2)^2}\right] + \mathcal{O}(C_3^2), \\
  p_{32} & = 
  \frac{3A_2\Omega_3^2\left[(A_3^2+2\Omega_3^2)^2+A_3^2(|C_3|^2+
    2A_3\text{Re}\,C_3)\right]}{(A_3^2 +
    2\Omega_3^2)\left[(A_3^2 +2\Omega_3^2)^2 + 3A_3^2(|C_3|^2+2A_3\text{Re}\,
    C_3)\right] + 2A_3^2\left[|C_3|^2|A_3+C_3|^2
    +(A_3^2+2A_3\text{Re}\,C_3)^2\right]}
  \nonumber \\
  & = \frac{3A_2\Omega_3^2}{A_3^2+2\Omega_3^2}\left[1
    - 4\text{Re}\,C_3\frac{A_3^3}{(A_3^2 +
      2\Omega_3^2)^2}\right] + \mathcal{O}(C_3^2).
\end{align}
\end{widetext}
\end{subequations}
Compared to two D systems the transition rates show the same
behavior as explained above for the three V systems. 
This is not surprising
as the quasisteady states are identical and as the D systems also
only interact via $C_3$ when they are in a light period.
Fig.~\ref{3Dp32} shows the exact transition rate $p_{32}$ compared to
the interaction free case. For distances of about a wavelength,
$p_{32}$ deviates up to $75\,\%$ from the rate without interaction. The first
peak at about $0.7$ wavelengths even reaches a maximum of seven times
the rate for independent atoms. For such small distances, however, one
would have to check the validity of the model (namely, that in a
particular intensity period most of the population is in a specific
subspace). Also one must keep in mind that all the experiments cited
here were performed at greater ion distances. 

\section{double and triple jump rate}
\label{doublejump}
The physical quantity investigated in the experiments of
Refs.~\cite{SaBlNeTo:86,Sa:86,ThBaDhSeWi:92,ItBeWi:88,BeRaTa:03} is the
double jump rate. This is the rate at which jumps between periods of
intensities that differ by twice the intensity of a single system
occur within a small time interval. In Ref.~\cite{AdBeDaHe:01} the double jump rate has been
expressed in terms of the transition rates $p_{ij}$ for two
dipole-interacting V systems. The same will be done here for three
systems.
As one can calculate directly from Eqs.~(\ref{pijcalc}) and
(\ref{pij}) there are no direct double jumps (i.e., $p_{ij}=0$ for
$|i-j|>1$). A double jump is
therefore defined as two successive jumps in the same direction which
occur within a time which is smaller than a time window $T_{\text{m}}$
so that they cannot be resolved. 
As there are four periods of different intensity in the fluorescence
of three three-level systems, there are also four different
possibilities for double jumps: From intensity zero to double
intensity, from single intensity to threefold intensity, and vice
versa. Therefore the whole double jump rate $n_{\text{DJ}}$ is the sum
of rates for the four different possible double jumps,
\begin{equation}
\label{nDJ1}
n_{\text{DJ}}=n_{\text{DJ}}^{20} + n_{\text{DJ}}^{31} +
n_{\text{DJ}}^{13} + n_{\text{DJ}}^{02}~.
\end{equation}
We first derive the rate for jumps from zero to double intensity. Each
period of zero intensity ends with one of single
intensity. The probability that the latter period is shorter than
$T_{\text{m}}$ is given by
\[
p_{T_1<T_{\text{m}}}=1-e^{-(p_{10}+p_{12})T_{\text{m}}}~.
\] 
\begin{figure}[b]
  \psfrag{nDJ}{$n_{\text{DJ}} [\text{s}^{-1}]$}
   \psfrag{r/lambda3}{$r [\lambda_3]$}
    \epsfig{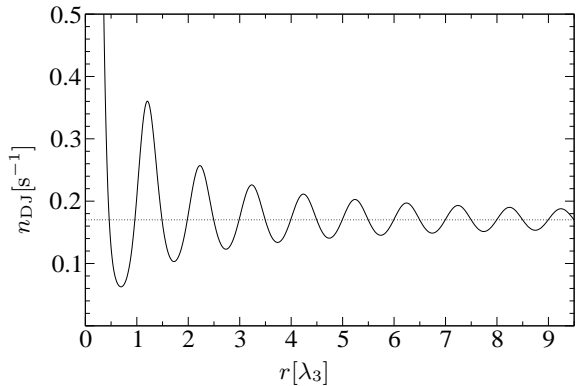}
   \caption{\label{3VnDJ}  Double jump rate $n_{\text{DJ}}$ for three
     dipole-interacting V systems. Solid line: $n_{\text{DJ}}$ up to second
     order in $C_3$. Dotted line: independent systems. Time window
     $T_{\text{m}} = 10^{-3}\,\text{s}$. Other parameter values as in Fig.~\ref{3Vp32}.}
\end{figure}
The branching ratio for the following period to be of double intensity
is $p_{12}/(p_{10}+p_{12})$. With the mean number of intensity periods
$I_i$ per unit time denoted by $n_i$ the rate $n_{\text{DJ}}^{02}$ is given by
\begin{equation}
n_{\text{DJ}}^{02}=n_0\frac{p_{12}}{p_{10}+p_{12}}\left(1-e^{-(p_{10}+p_{12})
  T_{\text{m}}}\right).
\end{equation}
Analogously one finds
\begin{equation}
n_{\text{DJ}}^{31}=n_3\frac{p_{21}}{p_{21}+p_{23}}\left(1-e^{-(p_{21}+p_{23})
  T_{\text{m}}}\right).
\end{equation}
\begin{figure}[b]
  \psfrag{p32 [1e-6]}{$\quad n_{\text{DJ}} [10^{-6}\text{s}^{-1}]$}
  \psfrag{r/lambda3}{$r [\lambda_3]$}
   \epsfig{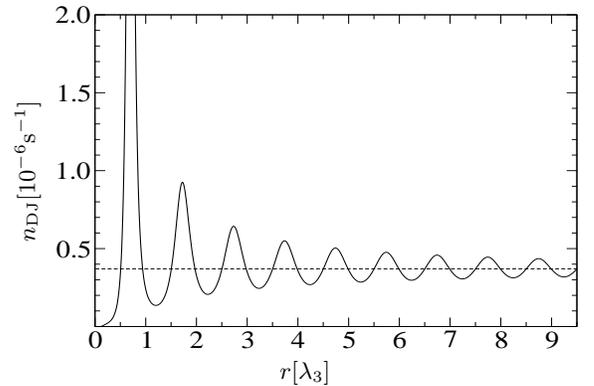}
   \caption{\label{3DnDJ}  Double jump rate $n_{\text{DJ}}$ for three
     dipole-interacting D systems. Dashed line: independent systems. 
     Time window $T_{\text{m}} = 5\cdot10^{-3}\,\text{s}$. Other parameter
     values as in Fig.~\ref{3Dp32}.}
\end{figure}
The remaining two rates are a little bit more complicated as the
periods of intensity $I_1$ and $I_2$ can be followed by a period with
either higher or lower intensity. The rates $n_{\text{DJ}}^{20}$ and
$n_{\text{DJ}}^{13}$ have thus to be supplemented with the branching
ratios $p_{21}/(p_{21}+p_{23})$ and $p_{12}/(p_{10}+p_{12})$,
respectively, yielding
\begin{equation}  
  n_{\text{DJ}}^{13}=n_1\frac{p_{12}}{p_{10}+p_{12}}\frac{p_{23}}{p_{21}+p_{23}} \left(1-e^{-(p_{21}+p_{23})T_{\text{m}}}\right)
\end{equation}
and
\begin{equation}
n_{\text{DJ}}^{20}=n_2\frac{p_{21}}{p_{21}+p_{23}}\frac{p_{10}}{p_{10}+p_{12}}\left(1-e^{-(p_{10}+p_{12})
  T_{\text{m}}}\right).
\end{equation}
Using the the relations
\begin{subequations}
\label{ni}
\begin{equation}
n_0=\frac{p_{10}}{p_{10}+p_{12}}n_1, \quad n_3=\frac{p_{23}}{p_{21}+p_{23}}n_2
\end{equation}
and
\begin{equation}
n_2=\frac{p_{12}}{p_{10}+p_{12}}n_1 + n_3, \quad n_1=n_0 +
\frac{p_{21}}{p_{21}+p_{23}}n_2
\end{equation}
\end{subequations}
the double jump rates can be simplified to
\begin{equation}
n_{\text{DJ}}^{02}=n_{\text{DJ}}^{20}=n_1\frac{p_{10}p_{12}}{(p_{10}+p_{12})^2}\left(1-e^{-(p_{10}+p_{12})
  T_{\text{m}}}\right)
\end{equation}
and
\begin{align}
n_{\text{DJ}}^{13}&= n_{\text{DJ}}^{31} \\*
& = n_1\frac{p_{12}p_{23}}{(p_{21}+p_{23})(p_{10}+p_{12})}\left(1-e^{-(p_{21}+p_{23})T_{\text{m}}}
\right).\nonumber
\end{align}
We denote the mean durations of the intensity periods by $T_i$ and
note that
\begin{equation}
\label{Ti}
T_0=\frac{1}{p_{01}}, \;\: T_1=\frac{1}{p_{10}+p_{12}}, \;\:
T_2=\frac{1}{p_{21}+p_{23}}, \;\: T_3=\frac{1}{p_{32}}.
\end{equation}
In addition they fulfill
\begin{equation}
\label{sumTi}
\sum_{i=0}^3 n_i T_i =1.
\end{equation}
The averaging window $T_{\text{m}}$ is much smaller than the mean
durations of the intensity periods. Therefore the exponential can be
expanded and with Eq.~(\ref{nDJ1}) one gets
\begin{equation}
n_{\text{DJ}}=2n_1\frac{p_{12}(p_{10}+p_{23})}{p_{10}+p_{12}}T_{\text{m}}.
\end{equation}
Using Eqs.~(\ref{ni}), (\ref{Ti}), and (\ref{sumTi}) we finally obtain
\begin{equation}
\label{nDJ2}
n_{\text{DJ}}=2\frac{p_{01}p_{21}p_{32}(p_{01}+p_{12})}{p_{21}p_{32}(p_{01}+p_{10})+p_{01}p_{12}(p_{23}+p_{32})}T_{\text{m}}
\end{equation}
as the double jump rate for three of either three-level
systems. 
\begin{figure}[t]
  \psfrag{nTJ[1e-3]}{$\quad n_{\text{TJ}}[10^{-3}\text{s}^{-1}]$}
  \psfrag{r/lambda3}{$r [\lambda_3]$}
   \epsfig{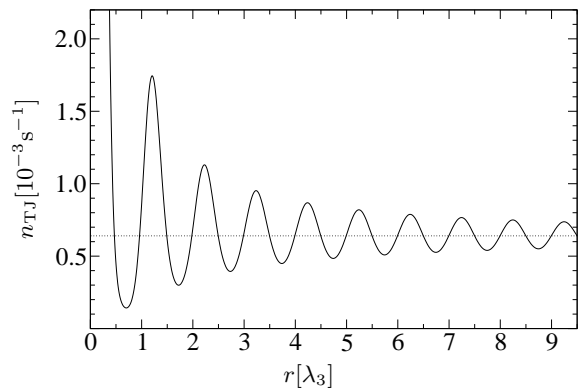}
   \caption{\label{3VnTJ}  Triple jump rate $n_{\text{TJ}}$ for three
     dipole-interacting V systems. Solid line: $n_{\text{TJ}}$ up to second
     order in $C_3$. Dotted line: independent systems. Parameter
     values as in Fig.~\ref{3VnDJ}.}
\end{figure}
A similar calculation yields for the triple jump rate
\begin{equation}
n_{\text{TJ}} =
2\frac{p_{01}p_{10}p_{12}p_{21}p_{23}p_{32}}{p_{21}p_{32}(p_{01}+p_{10})+ p_{01}p_{12}(p_{23}+p_{32})}T_{\text{m}}^2~.
\end{equation}
 Note that the defining time window $T_{\text{m}}$
enters quadratically in this case.
Figs.~\ref{3VnDJ} and \ref{3DnDJ} show plots of
$n_{\text{DJ}}$ for the V systems and the D systems, respectively,
whereas Figs.~\ref{3VnTJ} and \ref{3DnTJ} show plots of the triple
jump rate $n_{\text{TJ}}$ for both systems. For
the D systems the exact values for the $p_{ij}$ are used whereas for
the V systems only the expanded expressions up to second order in $C_3$
are used since $p_{23}$ and $p_{23}$ could not be calculated exactly
for the V systems. For the V systems there are cooperative effects of up
to $110\,\%$ for the double jump rate $n_{\text{DJ}}$ and $170\,\%$ for the
triple jump rate $n_{\text{TJ}}$ for distances of somewhat more than a
wavelength of the strong transition. For the same distance range the D
system shows cooperative effects of up to $150\,\%$ for both
$n_{\text{DJ}}$ and $n_{\text{TJ}}$. The first peak at three quarters
of a wavelength reaches 16 times the value for independent systems for
both rates.
\begin{figure}[t]
  \psfrag{nTJ[1e-11]}{$\quad n_{\text{TJ}} [10^{-11}\text{s}^{-1}]$}
  \psfrag{r/lambda3}{$r [\lambda_3]$}
   \epsfig{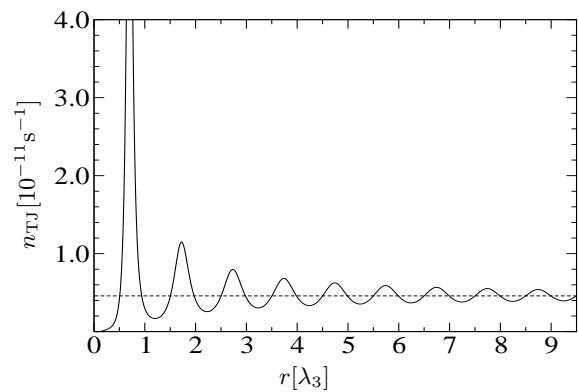}
   \caption{\label{3DnTJ}  Triple jump rate $n_{\text{TJ}}$ for three
     dipole-interacting D systems. Dashed line: independent
     systems. Parameter values as in Fig.~\ref{3DnDJ}.}
\end{figure}
For distances of about 10 wavelengths  cooperative effects of $15\,\%$ are still present for both systems.
In the case of the D system, which models the level configuration
of the  Hg$^+$ ions used in the experiments of
Refs.~\cite{ItBeWi:88,ItBeHuWi:87}, large cooperative effects only appear if
the Rabi frequency $\Omega_3$ is smaller than the Einstein coefficient
$A_3$. So, for the experimental parameters (i.e.,~$\Omega_3 > A_3$ and
$r/\lambda_3 \approx 15$) the effects are negligible, in agreement
with the experimental results.

\section{Conclusions}
We have investigated the effect of the dipole-dipole interaction on
three three-level systems showing macroscopic light and dark periods
in their fluorescence. This was done for the V and the D configuration,
respectively. The latter models the effective level configuration of
the Hg$^+$ ions in the experiments of Refs.~\cite{ItBeWi:88,ItBeHuWi:87}.
We have explicitly calculated the transition rates between the
different intensity periods for both configurations. In addition, the
double and 
triple jump rates have been derived from these transition rates.
Both systems show the same first order dependency on the coupling
parameter $C_3$, leading to an enhancement in the cooperative effects
by a factor of 2 for the transition rate $p_{32}$. This leads to
cooperative effects of about $100\,\%$ compared to the value for
independent systems for interatomic distances of somewhat more than a
wavelength of the strong transition. For the double and triple jump rates even
larger cooperative effects can be seen. For three D systems the first
peak at about three quarters of a wavelength is seven times higher
for $p_{32}$ and 16 times higher for $n_{\text{DJ}}$ and $n_{\text{TJ}}$ than
for independent atoms.

Although we did not treat the four-level system of Ref.~\cite{HaHe:03}
here, which models the Ba$^+$ ions of
Refs.~\cite{SaBlNeTo:86,Sa:86}, it is still possible to arrive at some
conclusions on this experiment from our results for the
three level systems. As was pointed out in Ref.~\cite{HaHe:03}, the
 results for two  D systems and two four-level systems are very
similar, in particular  in their first order term in $C_3$. It is therefore
very likely that the cooperative effects for three four-level systems are also
only enhanced by a factor of about 2, and since the effects for two
four-level systems were already negligibly small one can expect a similar
behavior also  for three of such systems.

\end{document}